\documentclass[aps,prb
]{revtex4-1}
\usepackage{amsmath,amsfonts,amssymb,bbold,bbm}
\usepackage[pdftex]{graphicx}

\begin{document}

	\author{
	Tetyana Ignatova$^\ddagger$$^*$,
	Sajedeh Pourianejad$^\ddagger$,
	Xinyi Li$^\dagger$,
	Kirby Schmidt$^\ddagger$,
	Frederick Aryeetey$^\vee $,
	Shyam Aravamudhan$^\vee $, 
	Slava V. Rotkin$^\dagger$$^\star$$^*$
}
\affiliation {~$^\ddagger$Department of Nanoscience, University of North Carolina at Greensboro, 2907 East Gate City Blvd, Greensboro, North Carolina 27401, USA; ~$^\dagger$Department of Engineering Science and Mechanics, The Pennsylvania State University, University Park, Pennsylvania 16802, USA; ~$^\vee$Department of Nanoengineering, North Carolina A$\&$T State University, 2907 East Gate City Blvd, Greensboro, NC 27401, USA; ~$^\star$Materials Research Institute, The Pennsylvania State University, Millennium Science Complex, University Park, Pennsylvania 16802, USA; ~$^*$corresponding authors. 
}

	\title{Multidimensional imaging reveals mechanisms controlling label-free biosensing in vertical 2DM-heterostructures}



	\begin{abstract}
	
Two-dimensional materials and their van der Waals heterostructures enable a large range of applications, including label-free biosensing. Lattice mismatch and work function difference in the heterostructure material result in strain and charge transfer, often varying  at nanometer scale, that influence device performance. In this work, a multidimensional optical imaging technique is developed in order to map sub-diffractional distributions for doping and strain and understand the role of those for modulation of electronic properties of the  material. As an example, vertical heterostructure  comprised of monolayer graphene and single layer flakes of transition metal dichalcogenide MoS$_2$ is fabricated and used for biosensing. Herein, an optical label-free detection of doxorubicin, a common cancer drug, is reported via three independent optical detection channels (photoluminescence shift, Raman shift and Graphene Enhanced Raman Scattering). Non-uniform broadening of components of multimodal signal correlates with the statistical distribution of local optical properties of the heterostructure. Multidimensional nanoscale imaging allows one to reveal the physical origin for such a local response and propose the best strategy for mitigation of materials variability and future device fabrication.

\end{abstract} 

	\maketitle

	\section{Introduction}

Emergent need to achieve better, more precise and sensitive drug detection in medicine and health care recently has been
addressed by developing biosensors based on two-dimensional materials (2DM)\cite{Kostarelos2014,Bolotsky2019,Zhu2019,Daus2021,Lee2014,Campuzano2017,Oh2021,Pang2020}. Not only 2D materials offer new response and/or transduction mechanisms and better performance, they can be used for label-free biosensing.  Importantly, 2DMs could be designed and/or integrated to generate several signals in response to a single analyte, as it will be illustrated below, or to respond by several channels to a group of substances in parallel, thus achieving a multimodal detection. 

The multimodal operation exceeds single-mode biosensing through its higher throughput as well as ability to differentiate the analyte from background signals in a complex media, and potentially allows the multiplexing of  biosensing\cite{Yen2015,Wang2015,Lee2016,Lei2019,Zhang2017}, {\em i.e.}, determining multiple analytes through a single test. While significant attention has been paid to exploring new 2D materials and demonstrating their biosensing capabilities at the level of single devices \cite{Zhang2015,Wang2014,ArjmandiTash2016,Sekhon2021}, overall knowledge on what allows successful multimodal detection and what limits biosensing capabilities of 2DM heterostructures is scarce. Atomically thin 2D materials, having an ultimate surface-to-volume ratio, may possess surface non-uniformities at the nanometer scale (atomic impurities/adsorbates/defects, wrinkles/ruptures) that modulate their  optical properties, although their importance and explicit role in producing material's variability yet to be studied. To a large extent, the difficulty to determine physical mechanisms that control performance of 2DM devices is due to disparate scales for atomic non-uniformities compared to a micrometer, or larger, size of active elements of a biosensor. Structural characterization with a high spatial resolution, such as electron microscopy, often does not detect materials optical properties, while optical microscopy lacks the required resolution. Thus, in order to reveal such mechanisms, multiple characterization tools should be combined and correlated\cite{Kolesnichenko2021}.  In this work, correlated multidimensional imaging, including Raman and near-field microscopies, scanning probe  and electron microscopies, was applied to unveil physical processes behind label-free multimodal detection of doxorubicin (DOX), an anthracycline cancer drug, by 2DM vertical heterostructures.

\begin{figure*}[b]
	\centering 
	\includegraphics[width=1
	\textwidth]{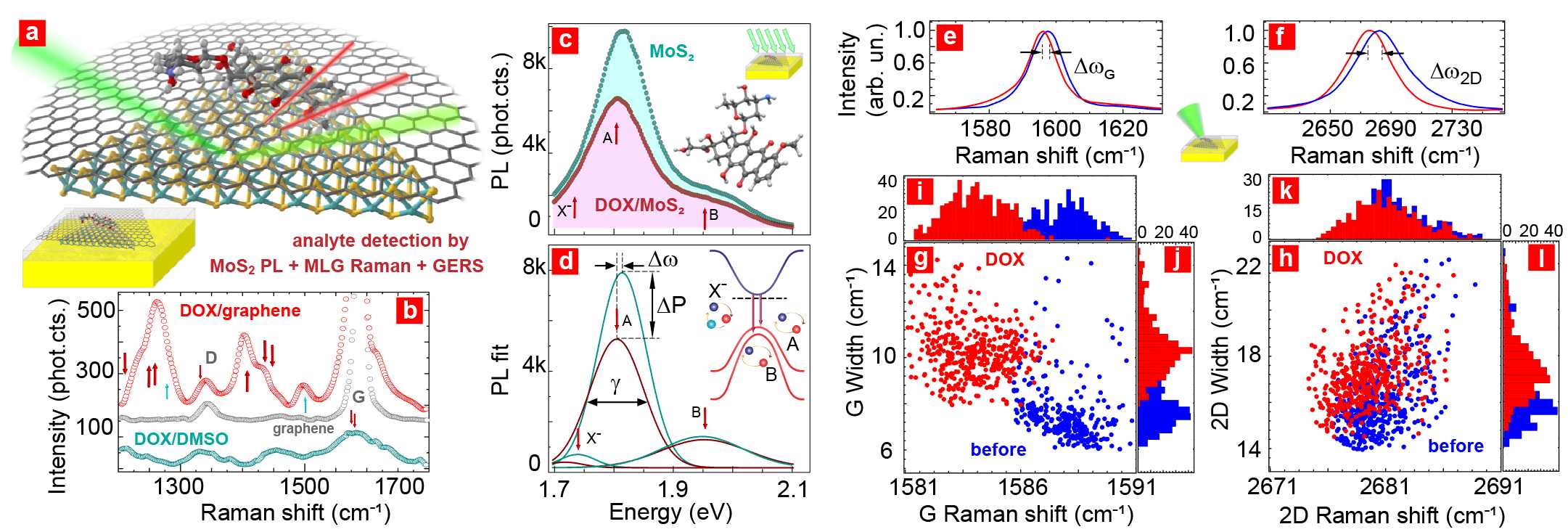}
	\caption{Multiplexed detection of doxorubicin drug. (a) Schematics of multimode detection by the combination of  MoS$_2$ photoluminescence, DOX GERS, and Raman shift of monolayer graphene. (b) GERS signal of DOX/MLG (red), vs. reference Raman spectra of DOX/DMSO solution (cyan) and MLG (gray); red (cyan) arrows mark DOX (DMSO) lines. (c) Modulation of MoS$_2$ PL spectrum: with DOX (red) and w/o DOX (cyan); inset shows DOX molecular structure. (d) Fitting of measured PL spectra from (c):  A/B-exciton and trion (X$^-$) lines are shown; modulation of peak position ($\Delta\omega$) and intensity ($\Delta P$) are indicated using A-exciton fit; inset shows the schematics of optical subbands of MoS$_2$. (e-f) Typical Raman spectra of MLG: with DOX (red) and before incubation (blue); G- and 2D-line intensities were normalized to unity. (g-h) Correlation plots and (i-l) partial distribution functions for peak position and width for G- and 2D-lines, measured locally, at diffraction limited spots across the sample; same color code as in (e-f); clear line red-shift and broadening are detected with DOX.}
	\label{fig:fig1}
\end{figure*}

Doxorubicin is one of the most common drugs against different types of cancer (haematological, thyroid, breast, ovarian, lung and liver cancer)\cite{Norouzi2020,Chen2021a,Zhong2017,He2020,Yuan2021}. Since DOX is known for certain drug resistance and side effects\cite{Carvalho2014,HOFMAN2015168,MITRY201617,Umsumarng2015}, an efficient and sensitive detection of the amount of DOX in various types of biological samples, potentially at the point-of-care, has significant value. Recently, DOX has been loaded on graphene oxide and other nanocomposites\cite{Sun2008,Chekin2019,Hasanzadeh2016,Yang2020,Pei2020}. Regular Raman microscopy, as well as surface enhanced Raman spectroscopy (SERS) were used to detect DOX in various  cell lines and real samples\cite{Zong2018,Gautier2013,Huang2013,Farhane2015,Farhane2017,Litti2016}. Here, optical signaling of the presence of DOX (deposited from solution) is demonstrated via three independent channels:  (1) graphene enhanced Raman spectra (GERS) of DOX, (2) Raman shift of monolayer graphene (MLG) and (3) photoluminescence (PL) shift of single layer MoS$_2$ (Fig.\ref{fig:fig1}).

Currently, two major approaches are implemented in biosensor technology: label-free and label-based sensing. While the latter shows high selectivity limited only by our ability to find a high-optical-contrast receptor with best binding to known analyte, the former is much more versatile, especially in terms of sensing a wide range of analytes, enabling agnostic biosensing, and being capable to detect yet unknown biothreats for which the receptors have not been developed. Though very promising, label-free biosensors require additional calibration due to lower specificity. To solve the problem sensing multiplexing, combined with machine learning, has been applied\cite{Zhang,Misun2016,Cui2020}.

\begin{table*}
	\caption{The PL fit parameters for Fig.\ref{fig:fig1}(d): upper/lower row corresponds to PL with/without DOX}
	\label{table:tablePL}
	\centering
\resizebox{\textwidth}{!}{	
	\begin{tabular}{|*{9}{c|}}
		\hline
		\multicolumn{3}{|c}{Trion} & \multicolumn{3}{|c}{A-exciton} & \multicolumn{3}{|c|}{B-exciton} \\ \hline 
		$\omega_c$, eV & $\gamma$, meV & P, cts. &$\omega_c$, eV  & $\gamma$, meV & P, cts.  &$\omega_c$, eV  &  $\gamma$, meV & P, cts.  \\ \hline
1.739 $\pm$ 0.002 & 60. $\pm$ 3.  & 32 $\pm$ 4 & 1.815 $\pm$ 0.0002 & 82.0 $\pm$ 0.3 & 793 $\pm$ 4 & 1.953 $\pm$ 0.001 & 135.8 $\pm$ 2. & 203 $\pm$ 1  
		\\ \hline
1.719 $\pm$ 0.003 & 60. $\pm$ 9.  & 15 $\pm$ 3 & 1.806 $\pm$ 0.002 & 90.7 $\pm$ 0.3 & 586 $\pm$ 3 & 1.955 $\pm$ 0.002 & 135.0 $\pm$ 2. & 197 $\pm$ 2  
\\ \hline
\end{tabular}
}
\end{table*}

In order to achieve multiplexed detection, arrays of different sensors could be integrated in one device\cite{Kalmykov2019}. To avoid unnecessary complexity of integration, mutimodal sensing materials and heterostructures are developed\cite{Novoselov2016,Jeong2015,Ma2020,AlaguVibisha2020}. Here we demonstrate multiplexed detection of doxorubicin by vertical heterostructure of monolayer graphene/transition metal dichalcogenide (TMDC) by measuring response of 2D materials in 3 optical channels: MoS$_2$ photoluminescence, graphene Raman shift and graphene enhanced Raman scattering of molecular fingerprint modes of the molecule itself.

\section{Results and discussion}
\subsection*{Label-free detection of Doxorubicin}

Molybdenum disulfide, a typical TMDC 2D material, is known to show strong PL signal\cite{Mak2013} which can be modulated by adsorbtion of molecular species\cite{Mouri2013,CatalanGomez2020,Aryeetey2021,Barja2019,Mitterreiter2021,Schuler2020,Thiruraman2018}. Fig.\ref{fig:fig1}(c) shows a profound change in PL spectrum of MoS$_2$ photoluminescence (PL) after incubation to 172 nM solution of DOX for 15 minutes (the large area integrated PL is presented here; to not be confused with local micro-PL discussed below). In order to understand physical mechanisms resulting in the DOX recognition, the PL band is fitted with individual excitation lines: as shown in the inset of  Fig.\ref{fig:fig1}(d), the MoS$_2$ optical transitions include typical B- and A-exciton subbands, trion (X$^-$) and, often, additional localized modes. Here the shift in mode peak position ($\Delta\omega$), peak intensity ($\Delta P$) and width ($\Delta\gamma$) are indicative for analyte absorption, resulted in subsequent charge transfer/doping and strain imposed in the 2D material. These shifts are specific for an analyte: panel (d) and data in Table \ref{table:tablePL} provide the values for DOX analyte. While upper B-exciton is barely influenced by the drug molecules (a small intensity difference is detected, see red arrow in panel (d)), both A-exciton and trion are red-shifted, have lower intensity and larger peak width, that all together lead to the spectral differences in panel (c). The ability to detect DOX at a low (sub-nM) concentration (and differentiate it from other components of a complex solution) would depend on amount of signal over the noise for the biosensor. Importantly, the variation of the signal in the pristine biosensing material adds to the total uncertainty and reduces the device performance as we discuss below.

\begin{figure}[!]
	\centering
	\includegraphics[width=0.4
	\textwidth]{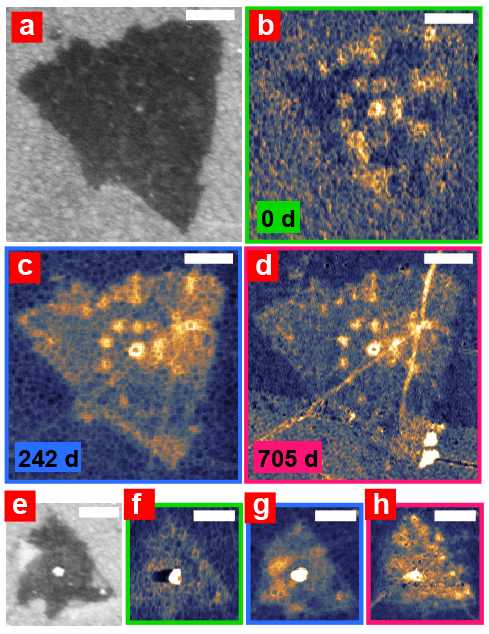}
	\caption{Stability test of  MoS$_2$/graphene vertical heterostructure. SEM (a,e) and sSNOM (b-d,f-h) images of two MoS$_2$ islands, randomly selected, coated with monolayer graphene. The island (a) shows nearly zero degradation after 242 days in ambient -- from (b) to (c), neither after 705 days -- from (b) to (d); the island (e) was selected near a tear in MLG and shows (g) partial oxidation near the central micro-crystallite of molybdenum after 242 days, followed by (h) almost complete oxidation of MoS$_2$ surface after 705 days. All scale bars are 1 $\mu$m.
	}
	\label{fig:fig12}
\end{figure}

\begin{table*}
	\caption{Measured GERS enhancement factors for major fingerprint Raman lines of DOX}
	\label{table:tableGERS}
	\centering
	\begin{tabular}{|*{8}{c|}}
		\hline
		Raman line position,	cm$^{-1}$ & 1236 &	1244 &	1260 &	1268 &	1326 &	1434 &	1613 \\
		\hline
		GERS enhancement factor & 6.4  & 7.0    &	23.3 &	23.3  &	1.8  &	2.9  &	2.1 \\
		\hline
	\end{tabular}
\end{table*}


Agnostic detection of a chemical or biothreat requires multiplexing the receptor signal  with additional channels, as there is no calibrated negative control for unknown analyte. In order to differentiate the signal from DOX against any other molecule potentially causing PL modulation, we measure the characteristic fingerprint Raman spectrum of DOX. Fig.\ref{fig:fig1}(b) shows the Raman spectrum of DOX/DMSO solution (cyan curve). However, DOX Raman lines (highlighted by red arrows) are mixed, superimposed and even obscured with DMSO (background) response (cyan arrows). Furthermore, the line intensity of analyte would be comparable to background even at a relatively high DOX concentration. On contrary, when deposited on graphene surface, most of DOX lines become clearly visible,  due to a significant GERS enhancement of the Raman signal of DOX (compare red and cyan curves). Table \ref{table:tableGERS} summarizes the amount of signal enhancement for particular lines. In our sample with only two substances, the intensity of fingerprint lines of DOX already allows to  confirm the analyte structure and determine the presence of analyte (which cannot be found from a PL data channel alone). While in general, for an agnostic biosensor, the whole Raman spectrum should be analyzed by machine learning correlation analysis of the data. Here, GERS, the second data channel, complements the PL detection which can provide information of the concentration of the drug (while the intensity of the GERS signal depends on enhancement factors and cannot be used to measure the amount of analyte). 


\begin{figure*}[!]
	\centering
	\includegraphics[width=1
	\textwidth]{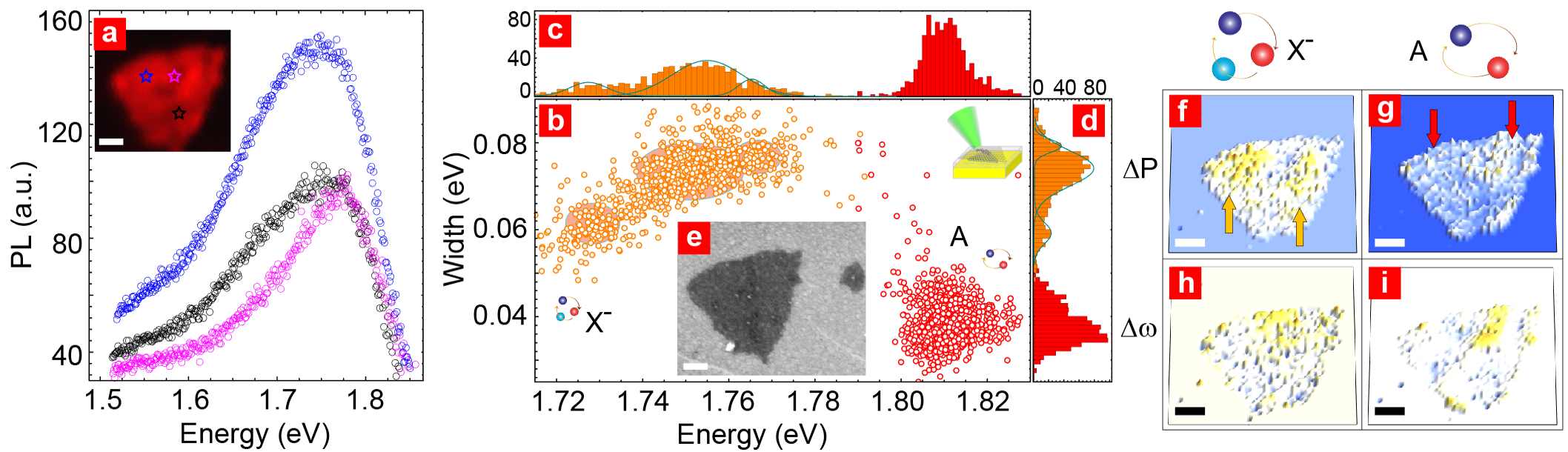}
	\caption{Local PL characterization of MLG/MoS$_2$ heterostructure. 
		(a) Single-point PL spectra of the MoS$_2$ island in (e). (inset) Total PL intensity map; stars show locations for the point spectra of the same color in main panel.
		(b) Correlation plots and (c-d) partial distribution functions for peak position and width for A-exciton (red) and trion (orange) lines, measured locally; several clusters are visible in trion data, highlighted by ovals in correlation plot and Gaussian envelope curves in distributions.
		(f-i) Confocal maps of MoS$_2$ PL: (top row) fitted intensity and (bottom row) peak position for (left) trion and (right) A-exciton; arrows show regions of higher PL intensity for trion (lower for A-exciton). 
		All scale bars are 1 $\mu$m.}
	\label{fig:fig2}
\end{figure*}

As Fig.\ref{fig:fig1}(b) shows, several DOX lines are superimposed with the Raman spectrum of graphene (gray curve corresponds to MLG reference), specifically with D- and G-lines near 1350 and 1600 cm$^{-1}$. While obscuring some of the DOX modes,  Raman spectra of graphene should be analyzed separately, yielding yet another channel, to be multiplexed with the PL and GERS data. Fig.\ref{fig:fig1}(e-f) shows pronounced red-shift and the width increase for two major lines of graphene, G- and 2D-band, upon interaction with the DOX analyte (red). Panels (g-l) show detailed statistical information on modulation of both line position and width for both modes; in contrast with previous optical data, each data point in this figure corresponds to a small local region on the sample, less than 0.1 $\mu$m$^2$, diffraction limited. Clearly, the data points aggregate in two separate clusters, though, point-to-point variability due to non-uniformity of the signal is non-negligible for 2D-mode (compare $\Delta\gamma/\Delta\omega$ correlation plot in panel (h) and partial distribution functions in panels (k-l)). Statistical distribution of the data from (g-l) contains important information about the  material/sample, which will be elaborated in detail next.

\subsection*{Stability of 2D van der Waals heterostructure materials}

Electron microscopy of  MoS$_2$/graphene vertical heterostructure, fabricated as described in Methods, reveals structural non-uniformities. A few typical images of several randomly selected single layer MoS$_2$ islands, coated with MLG, are shown in Fig.\ref{fig:fig12}(a,e) and  Fig.\ref{fig:fig2}(e). White nanocrystallites, likely made of insulating molybdenum oxide, charged under e-beam, are seen either in the center of the island (metal nucleation site) or at the edge (metal precipitation site); in some cases those grow to microcrystals of Mo$_2$O$_3$ (see Fig.\ref{fig:fig12}(e)) of characteristic triangular (or rectangular, not shown here) shape and size up to 1/2 micrometer. Graphene seems to be conformal to the substrate, making short wrinkles between nanoscale posts (10-20 nm tall).

While the surface of  MoS$_2$ islands appears mostly uniform in scanning electron microscopy (SEM) image, optical properties of 2DM demonstrate substantial variation in agreement with Raman and PL statistics  from  Fig.\ref{fig:fig1} and Fig.\ref{fig:fig2}. The variability of PL in pristine material could produce uncertainty in detection of the analyte. In order to find the origin for such a variation, scattering scanning near-field optical microscopy (sSNOM) has been applied. Careful alignment of large area scans of the same heterostructure allows us to correlate different characterization channels (including SEM, scanning probe imaging, as well as PL and Raman microscopy, having a lower resolution though). In Fig.\ref{fig:fig12}(b-d) the 
sSNOM image (2nd harmonic optical amplitude, see Methods for details) reveals variation of surface impedance of MLG/MoS$_2$ heterostructure at the sub-micrometer scale, not captured by SEM (or AFM). We argue that a series of bright regions (on the darker background of MoS$_2$) correspond to the local defects of the TMDC material. Indeed, we regularly observe such a contrast at the edge of the island which is known to be prone to partial oxidation. Similar regions in the bulk of the island should correspond to concentrated sulfur vacancies, reactive to oxygen, and formation of oxy-sulfate regions, often appearing as nanoscale posts that ruckle graphene around (a near-field phase image in  Fig.\ref{fig:fig4}(h) has the best contrast which allows to resolve nano-posts). The series of maps in Fig.\ref{fig:fig12}(b-d) and (f-h) show evolution of such regions protected (or non-protected) by graphene coating: the larger island (a), covered with intact MLG, preserves the same number of partially oxidized regions after nearly 2 years in ambient, except for a small oxide crystal grown in the bottom right corner, where a trench in graphene (dark line) opens an access to the air.  On contrary, the small island (e) has the MLG coating cracked; as a result, the surface is slowly oxidized over the course of retention period, almost entirely on the map in panel (h). sSNOM mapping also shows that the large graphene wrinkles (bright diagonal lines in panel (d)) do not lead to alteration of optical properties. On the opposite, the oxy-sulfate regions will be shown to generate non-uniform doping of the MoS$_2$ and (graphene), leading to the PL variability over the sample.



Local fluctuations of PL in the pristine material were analyzed in another island of the same 2DM vertical heterostructure mapped by SEM in Fig.\ref{fig:fig2}(e) and in  Fig.\ref{fig:fig3}(b) and Fig.\ref{fig:fig4}(h) by sSNOM. Several features are clearly resolved: graphene ruptures (not reaching the island), an oxide crystallite at the edge of the island, a few oxy-sulfate nano-posts and graphene wrinkles around the posts, and several regions of darker SEM contrast (likely, more conductive than bare MLG), potentially indicating doping/Fermi level variation. Confocal PL image of the same area is presented in Fig.\ref{fig:fig2}(a), inset. The large non-uniformity of PL intensity is followed by substantial variability of PL line shape (cf. the curves in main panel taken at three locations shown in the inset). Similar to the large area PL data in Fig.\ref{fig:fig1}(c), the main variability of micro-PL results from the A and X$^-$ states, to be analyzed separately. Panel (b) presents the correlation plot for fitted PL peak position and width for A-exciton (red) and trion (orange) states by the local optical probe on the surface of MLG/MoS$_2$ heterostructure shown in Fig.\ref{fig:fig2}(e). MicroPL reveals large non-uniformity in optical signal. Trion partial distribution functions for both  $\Delta\gamma$ and $\Delta\omega$  show 3 major clusters (highlighted by ovals in panel (b) and green curves in (c-d)), that correspond to the  regions of heterostructure where materials properties are locally modulated. 

Maps in panels (f-i) show actual distribution of the peak position, $\Delta\omega$, and peak intensity, $\Delta P$, with diffraction limited resolution. Importantly, the intensity maps show the anti-correlation for PL strength of A-exciton and trion (as indicated by red and orange arrows): the trion PL is the highest where the A-exciton PL is depressed, compare locations for trion-dominated (blue/black) and exciton-dominated (purple) PL curves in panel (a). Such a correlation may result from non-uniform doping of the  MoS$_2$ island. Indeed, in a highly-doped area the neutral excitons are bound to free charges and, thus, converted into trions\cite{Mouri2013}.


\subsection*{Multidimensional characterization of heterostructure materials}

\begin{figure}[!]
	\centering
	\includegraphics[width=0.45
	\textwidth]{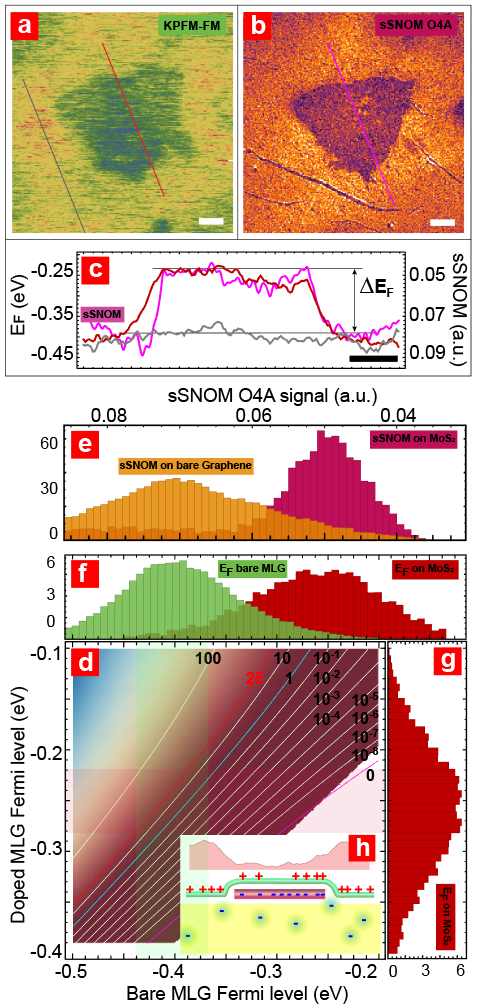}
	\caption{Correlation of MLG work function data with sSNOM optical surface impedance. Aligned maps for (a) KPFM and (b) sSNOM (4th harmonic) amplitude. (c) Cross section profiles across the  MoS$_2$ area (KPFM, red and sSNOM, pink) vs.MLG reference (KPFM, gray), taken along the lines of the same color in (a-b). (d) Calculated electron density in MoS$_2$ heterostructure, log-scale, vs. Fermi levels in bare/doped graphene off/on the island. (f-g) Partial distribution functions for measured $E_F$ in bare graphene (off island, green) and graphene  doped by the MoS$_2$ (on island, red) from KPFM map in (a).   (e) Partial distribution functions for sSNOM signal from (b) to calibrate near-field signal by  $E_F$. Note common abscissa axis for panels (d,f), not (e). Inset (h) shows schematics of charge transfer in the vertical heterojunction on SiO$_2$ substrate with negative charge traps. Pink curve outlines the variation of MLG work function.
		All scale bars are 1 $\mu$m.
		}
	\label{fig:fig3}
\end{figure}

Although useful to shed the light on the PL variability, the confocal PL characterization neither has enough spatial resolution nor enables assessing the MoS$_2$ doping level to uncover the mechanisms of non-uniform optical signaling. Instead, we developed a multidimensional imaging combining sSNOM and Kelvin probe force microscopy (KPFM) to be correlated with PL (and Raman) microscopy. In Fig.\ref{fig:fig3}(a-b) two maps of the same island -- using KPFM (work function) channel and sSNOM (optical surface impedance) channel -- show identical contrast, further detailed in panel (c) where the cross section profiles allow to quantify the variation of the Fermi level of graphene above the MoS$_2$ layer.   The profile of work function is schematically shown in  Fig.\ref{fig:fig3}(h). Charge transfer in the vertical heterojunction decreases the carrier density in both graphene and  MoS$_2$ underneath, thus, decreasing the magnitude of graphene work function and doping level. The KPFM probe is in contact with the outermost layer of the heterostructure, graphene, thus it measures the work function of MLG. Graphene above the island appears negatively doped by MoS$_2$. The MLG Fermi level, taken with respect to graphene Dirac point, is negative, corresponding to p-doping.  Statistical distribution of the Fermi level values of graphene on/off the island is shown in panel (f) by red/green histogram. Knowing  $E_F$ in bare graphene and in the vertical heterostructure allows us to calculate the MoS$_2$ doping level, Fig.\ref{fig:fig3}(d). Using median values for $E_F$, it can be estimated to lie in the range $1-25\times10^{12}$~cm$^{-2}$, which is further corroborated by independent Raman data below.

Comparison of KPFM and sSNOM profiles in Fig.\ref{fig:fig3}(c), as well as the distribution functions in Fig.\ref{fig:fig3}(f-e), allows to calibrate the near-field signal in terms of the Fermi level of the heterostructure. Then, one could interpolate the charge transfer/doping data to the nanometer features, only resolved by sSNOM (such as wrinkles, oxy-sulfate regions, etc.), and thus, determine the origin for PL non-uniformity. 

Enhanced resolution of sSNOM allows us to determine 5 sources of non-uniform doping in the vertical van der Waals heterostructures as shown schematically in Fig.\ref{fig:fig3}(h). (i) The primary doping is defined by conditions of the MoS$_2$ synthesis: it is known that often the stoichiometry of TMDC is slightly off the equilibrium values. Deficiency in sulfur leads to creation of surface vacancies, typically resulting in n-doping\cite{Komsa2015}. (ii) Filling of the S-vacancy with oxygen or CH-group yields weaker n- or p-doping\cite{Mouri2013,Zheng2020}, which was shown to be localized near the defect site\cite{Nan2014}. (iii) In Mo-abundant synthesis, small micro-crystallites of metal molybdenum form, later oxidized to MoO$_x$, or forming MoO$_x$S$_y$ domains. (iv) In the heterostructure, work function and/or Fermi level difference between the layers results in charge transfer between the layers. Typically p-doped MLG would become an acceptor for electrons transferred from n-doped MoS$_2$. Finally, (v) the Si/SiO$_2$ substrate supports the heterostructure, which is known to have a high density of traps at the interface. Such traps, if charged, produce a substantial field and shift of the Fermi level in all 2DM layers above it, generating a random Coulomb potential for charge carriers both in  MoS$_2$ and graphene. 

\begin{figure*}[!]
	\centering
	\includegraphics[width=1
	\textwidth]{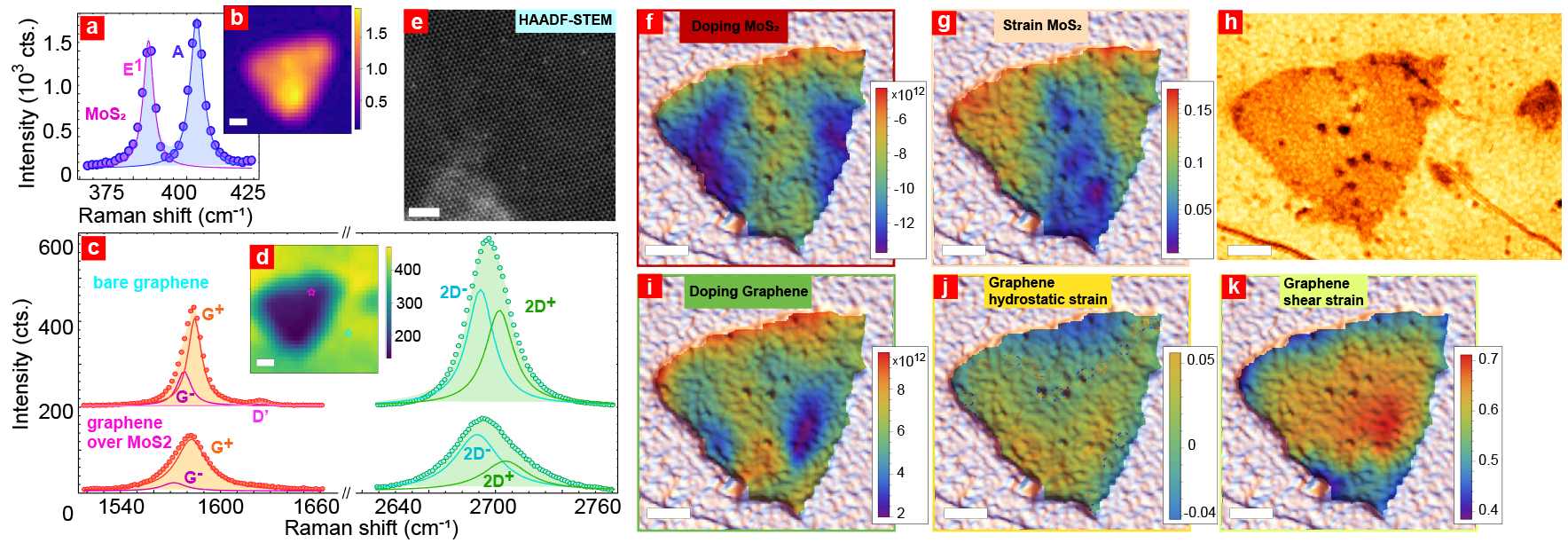}
	\caption{Raman mapping of doping and strain non-uniformity in the heterostructure. (a) Typical MoS$_2$ Raman spectrum, fitted with E$^1$ and A-lines. (b) A-line intensity map. (c) Typical Raman spectra for MLG off/on  MoS$_2$ island, fitted by G (orange), D' (pink) and 2D (green) lines; splitting of G- and 2D-lines is shown in the fit. (d) Raman map of 2D-amplitude showing the island location, cf. map in (b).	
		(e) HAADF-STEM image of  MoS$_2$ lattice: notice grain boundaries and individual defects; scale bar is 2 nm. 
(f,g) Calculated  doping and strain for MoS$_2$ layer overlaid with SEM map; (h) sSNOM phase image of the same area; (i-k) MLG doping, hydrostatic and shear strain maps.
		All scale bars, except in (e), are 1 $\mu$m.
	}
	\label{fig:fig4} 
\end{figure*}

Additional evidence for the existence of defects/vacancies in TMDC lattice is provided by high-angle annular dark-field (HAADF) Scanning Transmission Electron Microscopy (STEM) imaging.  Fig.\ref{fig:fig4}(e) shows atomic resolution map of a typical MoS$_2$ island. The boundary between dark area and the lower contrast area likely reflects the grain boundary which separates regions of different lattice orientation. Such a twin boundary produces strain and may result in localization of electronic states. Furthermore, several (3-fold) individual defects are seen in the STEM image (approximately half a dozen per 200 nm$^2$ which corresponds to ca. $3\times10^{12}$~cm$^{-2}$).

Multiple sources of optical non-uniformity, stemming from the variation of the doping level, have been further studied with micro-Raman imaging: typical Raman spectra of  MLG/MoS$_2$ heterostructure are shown in Fig.\ref{fig:fig4}(a,c). 
Panel (a) presents A- and E$^1$-modes of MoS$_2$ layer, A-intensity map is shown in inset (b). Mode frequencies, fitted as in (a), allow to determine the strain and doping\cite{Rao2019} of the island underneath the graphene, see Methods, generating the maps presented in panels (f,g). Consistent with the KPFM data, Fig.\ref{fig:fig3}(a),  MoS$_2$ doping is lower along the vertical axis of the island, thus, both the amount of charge transfer and graphene $E_F$ should be lower. 
Charge doping and strain in graphene have been calculated using a similar procedure\cite{Neumann2015,Mueller2017}. 
Upper/lower curves in panel (c) correspond to MLG Raman lines off/on TMDC, where the location of the island is clearly seen, {\em e.g.}, in the map of 2D-amplitude (d).  Fig.\ref{fig:fig4}(i,j) show graphene doping and isotropic/hydrostatic strain. Furthermore, the splitting of the G- and 2D-doublet modes (see the fitted curves in panel (c)) yields\cite{Narula2012} the shear (non-isotropic) component of the strain, panel (k).

High-resolution map reveals that the hole carrier density in graphene increases next to the location of a large MoO$_x$ crystallite, which should indicate additional chemical doping. Besides doping, all nanoscale features of heterostructure morphology make contributions to the uniform and non-uniform components of graphene strain, thus making Raman line width larger than the natural width\cite{Neumann2015}, due to the statistical broadening.


\subsection{Outline}
\label{sec:conclusions} 

Cumulatively, multidimensional characterization data above revealed existence of non-uniformities in 2D materials at the nanoscale and allowed to identify  doping and/or strain variations as the origin of statistical distribution of the optical signals used in all three recognition channels (PL shift, Raman spectroscopy and GERS). When integrated over the device area, such a variability in local response would translate in a broadening of the biosensing spectral signal, thus, raising device-to-device variability and, ultimately, lowering the sensitivity and the limit-of-detection by increasing background and/or systematic error. While the variability of individual device response often could be addressed by careful calibration against known analytes, such a fluctuation and spread of the integrated response would affect biosensing accuracy and, certainly, reduce the ability to perform precise biosensing in the agnostic detection mode. Presented study suggests that in order to improve the performance of biosensors based on 2DM heterostructures, non-uniformity of doping and strain -- two major mechanisms for optical signal variation -- must be addressed. Currently, most of 2DM heterostructures are fabricated by transfer methods, that are known to produce both strain and doping\cite{Leong2019,Bousige2017,Banszerus2017} (especially for wet transfer). New methods of strain-free and doping-free transfer need to be developed\cite{Leong2019,Seo2021}. Alternatively, such heterostructure materials should be fabricated in-situ, in synthetic facility, to preserve the layer epitaxy and exclude contamination between the layers.

\section*{Methods}

\subsection{Sample fabrication} 
The monolayer MoS$_2$ was grown on a Si substrate with 300 nm thick SiO$_2$ by Chemical Vapor Deposition (CVD) method as described in\cite{Aryeetey2021}. Optimization of synthesis parameters and stoichiometric ratio of molybdenum to sulfur resulted in producing triangular MoS$_2$ islands with low defect density (cf. STEM image in Fig.\ref{fig:fig4}(e)), predominantly single layers, with low surface coverage. Monolayer graphene was grown by CVD on Cu foil. MLG was transferred onto MoS$_2$ using the conventional PMMA assisted transfer technique\cite{Gao2012}. The SEM image of resulted heterostructure is shown on Fig.\ref{fig:fig2}(e).

\subsection{Sample Characterization.}
SEM sample imaging was performed in a field emission scanning electron microscope Zeiss Auriga FIB/FESEM. Atomic resolution images of  monolayer MoS$_2$ samples transferred onto Quantaifoil TEM grids were recorded using Nion Ultra HAADF-STEM operating at 60 kV with 3rd-generation C3/C5 aberration corrector and 0.5 nA current in atomic-size probe $\sim 1.0 - 1.1$\AA~ (NCATSU).
Confocal PL and Raman characterization were performed using a Horiba Jobin Yvon LabRAM HR-Evolution Raman system, 488 nm (for Raman) and 532 nm (for PL) laser excitation wavelengths were used; Horiba XploRA Raman system was used for taking Raman spectra at 532 nm of excitation. Analysis of PL and Raman characterization was performed using home-written codes. 
 
sSNOM maps were collected using scattering type scanning near-field optical microscope (custom-built Neaspec system) in pseudo-heterodyne mode (tapping amplitude $\sim$70 nm, ARROW-NCPt probes by Nanoworld $<$25 nm radius), excitation by CW Quantum Cascade Laser (MIRCat by Daylight) at power $<$ 2 mW in focal aperture at 1577-1579 cm$^{-1}$ (6.333-6.341 $\mu$m). Amplitude and phase of high order harmonics ($\ge 2$) are proportional to the local impedance of the sample under the tip.

The AFM/KPFM was performed using Dimension Icon AFM in PeakForce Kelvin Probe Force Microscopy in frequency modulated mode (PFKPFM-FM, Bruker Nano Inc., Santa Barbara, CA) utilizing a PFQNE-AL probe (Bruker SPM Probes, Camarillo, CA).  Prior to measuring the samples, the  KPFM response of the probe was checked against an Au-Si-Al standard and the work function of the Al reference metal layer was calibrated against a freshly cleaved highly oriented pyrolytic graphite (HOPG) reference sample (PFKPFM-SMPL, HOPG-12M, Bruker SPM Probes, Camarillo, CA); 4.6~eV was used for the work function reference value for HOPG.

\subsection{Supporting Information} \par 
Supporting Information is available from the Wiley Online Library or from the author.

\section*{Acknowledgments:} 
	
Authors are personally thankful to Drs. T. Tighe, T. Williams and M. Wetherington (MCL, PSU). S.V.R. acknowledges NSF support (CHE-2032582). T.I. and K.S. acknowledges NSF support (CHE-2032601). T.I. acknowledges Sample Grant from The Pennsylvania State University 2DCC-MIP
, which is supported by NSF cooperative agreement (DMR-1539916).  Work at PSU sSNOM facility has been partially supported by NSF MRSEC (DMR-2011839). Part of this work was performed at the Joint School of Nanoscience and Nanoengineering (JSNN), a member of the Southeastern Nanotechnology Infrastructure Corridor (SENIC) and National Nanotechnology Coordinated Infrastructure (NNCI), which is supported by the NSF grant (ECCS-1542174). Scanning Transmission Electron Microscope imaging was conducted at the Center for Nanophase Materials Sciences at ONRL, which is a DOE Office of Science User Facility. 



%
%
%

\section*{Supplementary Information}


\section{Strain and Doping analysis}
The background signal of Raman spectra for monolayer graphene (MLG) was fit and subtracted using air PLS \cite{Zhang2010}. Then peaks were fit for both MLG and TMDC spectra using the non-linear least-squares minimization and curve-fitting library (LMFIT) for python. Peaks were fit using Lorentzian line shapes around the D, G$^+$, G$^-$, 2D$^+$ and 2D$^-$ peaks, as well as around nearby shoulder peaks if they were distinguishable. 

\begin{figure*}[b]
	\centering 
	\includegraphics[width=.5 
	\columnwidth]{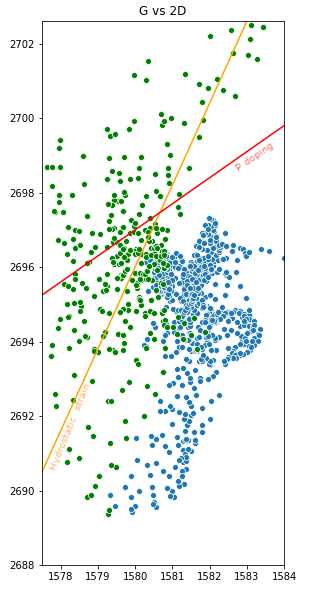}
	\caption{G vs 2D plot showing the split of graphene on the bare SiO$_2$ substrate (light blue) and over MoS$_2$ island (green). Red/yellow line indicates a characteristic slope for G-2D data correlation caused by pure doping/strain (isotropic biaxial).}
	\label{fig:SI-01}
\end{figure*}

The initial separation of strain and doping is accomplished by examining the central peak position of the 2D and G line fits. These Raman frequencies are sensitive to both strain and doping because of the change in lattice constants and force fields that effect the phonon frequencies. Lee et al.  created a procedure for extracting the strain and doping of graphene through the statistical analysis of the changes in the Raman frequency position \cite{Lee2012}. By plotting the 2D and G peaks against each other we are able to see trends in the spectra which represent modulation either by strain or by doping, or both. Strain is seen in the MLG Raman data as a cluster in of the 2D/G correlation plot (Figure \ref{fig:SI-01}) with a linear slope of approximately 2.2.  The linear slope for p-doped MLG is approximately 0.75 which is also seen in Figure \ref{fig:SI-01}. At very low values of p-doping and n-doping the dependence should be nonlinear, though, due to Fermi velocity (density of states) renormalization. 

It is known that graphene on MoS$_2$ and SiO$_2$ substrate is typically p-doped. Assuming linear correlation with Raman frequencies, we can extract the relative change in strain and doping by solving the linear equation system:
\begin{eqnarray}
	\left(\begin{array}{l}
		\omega_G
		\\ \\\omega_{2D}
	\end{array}\right)
	=
	\begin{array}{|ll|}
		a_{G,\varepsilon}\qquad & a_{G,\rho}
		\\ &\\
		a_{2D,\varepsilon}  & a_{2D,\rho}
	\end{array}
	\left(\begin{array}{l}
		\varepsilon
		\\ \\
		\rho
	\end{array}\right)
	\label{SI-01}
\end{eqnarray}
Where the vector ($\omega_{G}$,$\omega_{2D}$) should be calibrated against unstrained and undoped graphene reference sample. 

Graphene has two different polarizations of optical modes that are degenerate at zero strain. Depending on the axial direction of (uniaxial) strain, position of one of the modes shifts with respect to the other one. This generates a Raman doublet for general strain. Knowing a particular strain configuration is only possible with Raman mapping in polarized light, which resolved the polarization of a phonon mode.  However, even in the case of non-polarized Raman data, position of individual components of the doublet allows to separate the isotropic and anisotropic components of the strain. The latter corresponds to the shear strain, although in order to determine specific shear direction, a polarized spectroscopy will be required, also on a calibration sample with know lattice orientation.

\begin{figure*}[!]
	\centering 
	\includegraphics[width=1 
	\columnwidth]{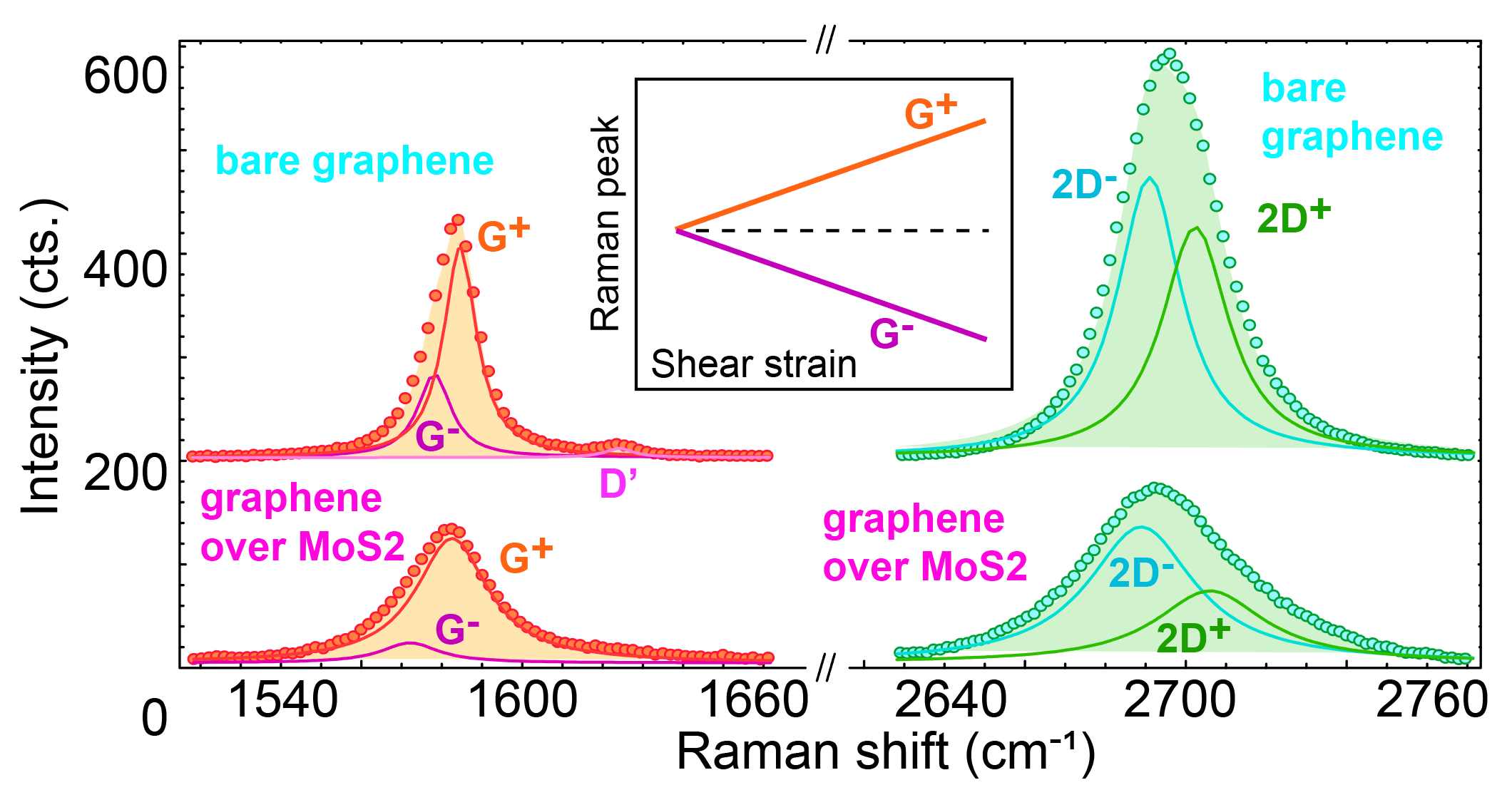}
	\caption{Main figure reproduces the data from Fig. 5c: the dots and filled line represent the experimental data and the total fitted curve. The individual components of the doublet are shown with the thin lines. Additional D'-component is needed for fitting the spectrum in vicinity of the G-doublet for bare graphene. (inset) Schematics of G-line splitting with the shear strain. Hydrostatic strain, on contrary, shifts the whole doublet but does not influence the splitting.}
	\label{fig:SI-02}
\end{figure*}

Mueller et al. developed a formalism to separate the doping and hydrostatic strain and shear strain components\cite{Mueller2017} . Critically, the shear strain component does not change the strain/doping correlation, that is, the slope of the curves in  Figure \ref{fig:SI-01}. While the hydrostatic component does not affect the splitting of the 2D or G peaks into doublet, as it shown on Figure \ref{fig:SI-02}. The amount of the splitting allows us to determine the magnitude of the shear strain, while the magnitude of the hydrostatic strain can be determined by examining the averaged peak position (after splitting). We can then determine the magnitude of the strain components and the doping by examining the shift of the peaks in a “zero strain” case or a “zero doping” case. Parametrization follows the paper by Das et al.: in the undoped case the 2D peak shifts at a rate of 1.04 cm$^{-1}$ per 10$^{12}$ cm$^{-2}$ hole density \cite{Das2009}. We use 2D splitting data and the Grueneisen parameter and the shear deformation potential from \cite{Mueller2017} to determine the strain components from:
\begin{equation}
	\omega_{2D}^{\pm}= \langle\omega_{2D}\rangle\, \left(-\alpha\,\varepsilon_h\pm \beta\,\varepsilon_s\right)
	\label{SI-02}
\end{equation}
where  $\alpha= 1.8$ is the Grueneisen parameter for MLG, and $\beta= 0.99$ is the shear deformation potential.

\begin{figure*}[!]
	\centering 
	\includegraphics[width=1 
	\columnwidth]{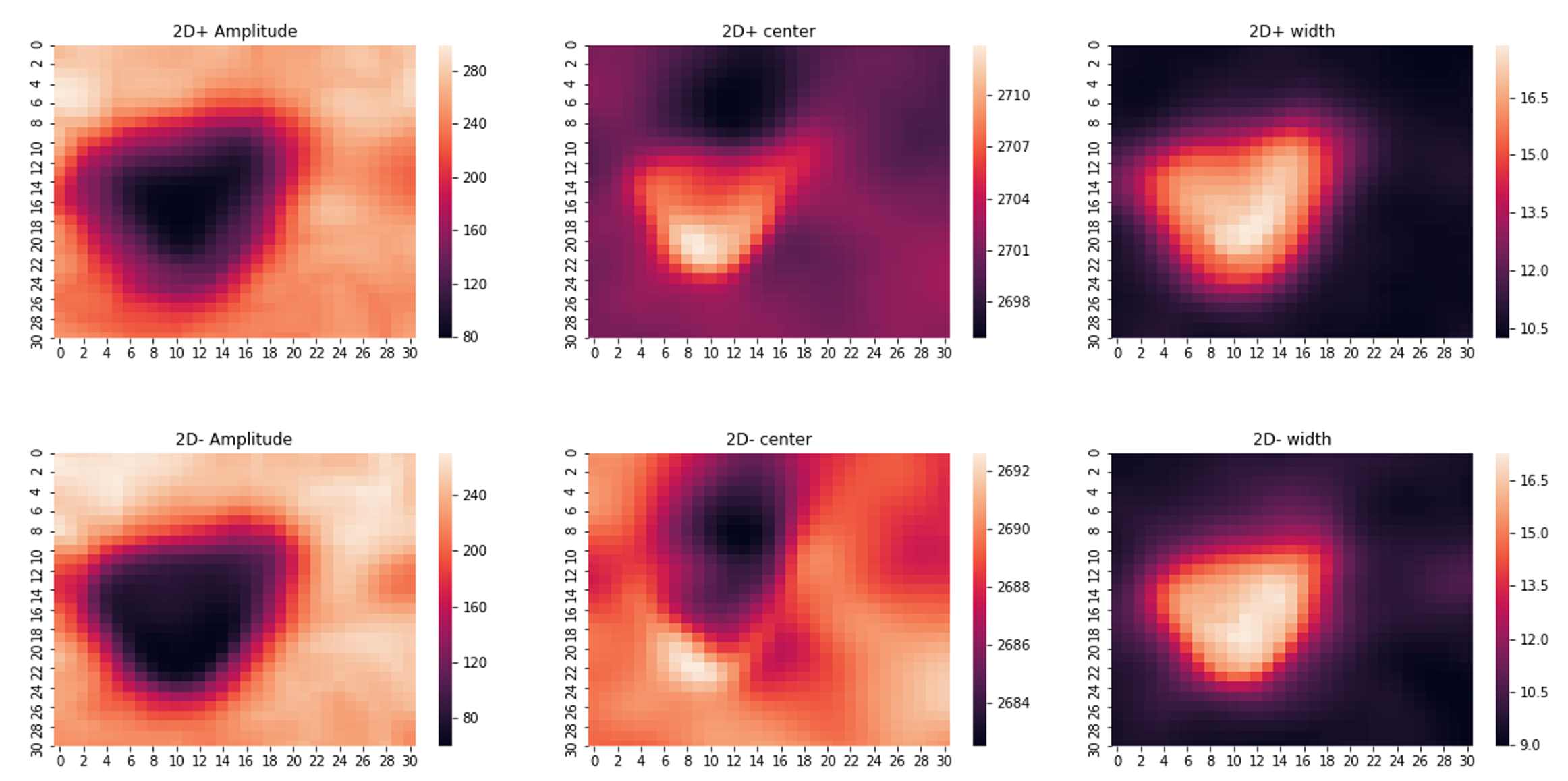}
	\caption{The maps showing the fitted parameters for splitting of 2D peaks.}
	\label{fig:SI-03A}
\end{figure*}

\begin{figure*}[!]
	\centering 
	\includegraphics[width=1 
	\columnwidth]{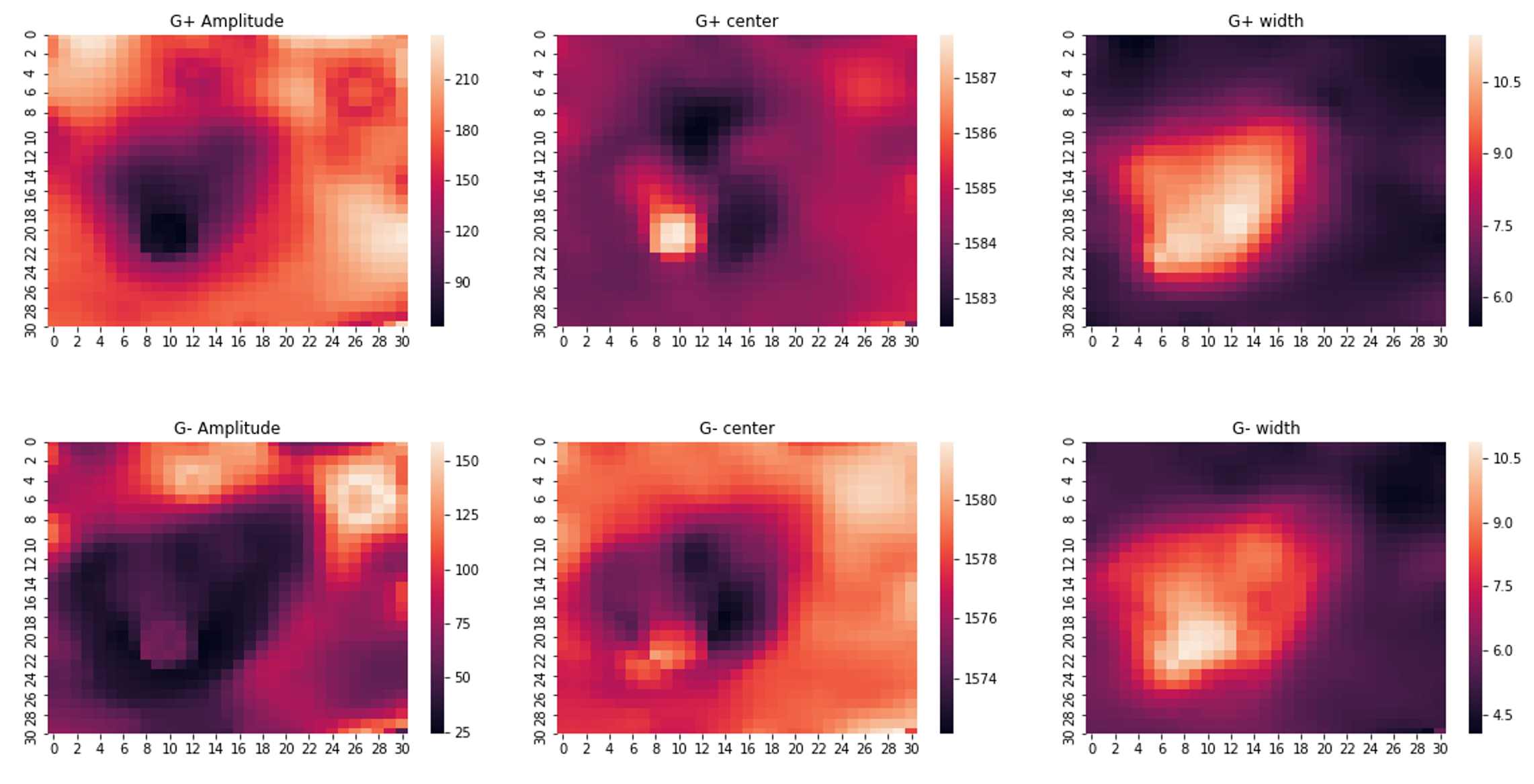}
	\caption{The maps showing the fitted parameters for splitting of G peaks.}
	\label{fig:SI-03B}
\end{figure*}

\begin{figure*}[!]
	\centering 
	\includegraphics[width=1 
	\columnwidth]{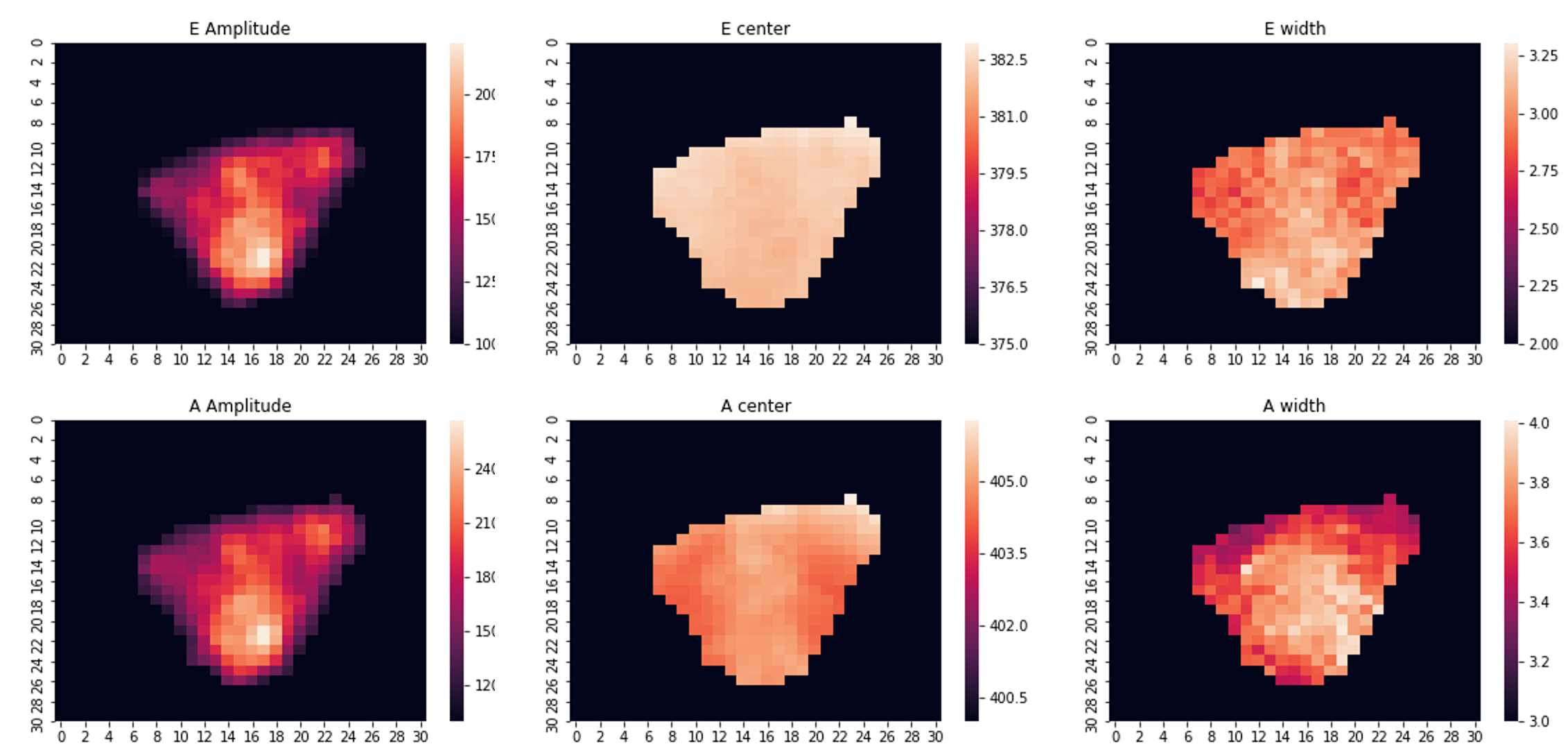}
	\caption{The maps showing the fitted parameters for MoS$_2$  peaks.}
	\label{fig:SI-04}
\end{figure*}

The strain and doping of MoS$_2$ can also be determined from Raman correlation data (Rao et al.2019). Peaks for MoS$_2$ were fit in the same way as the graphene peaks (Figure \ref{fig:SI-04}). For the case of MoS$_2$ we compare E peak and A peak that are near 382 and 404  cm$^{-1}$  respectively. The E peak position is more sensitive to strain, similar to the 2D peak of MLG, while the A peak position is more sensitive to doping, like the G peak  of MLG. The slope for strain correlation is $\sim$4; the slope for doping is $\sim$0.12. We can then use the undoped E peak position, and a Gruneisen parameter for MoS$_2$  of $\sim$0.86, to obtain the average strain. Then we examine the unstrained A peak position which shifts at a rate of 4 cm$^{-1}$ per 1.8~10$^{13}$ cm$^{-2}$ \cite{Chakraborty2012} and determine the doping. Unlike graphene, the peak splitting in MoS$_2$ is absent. 

\section{Characterization by Peak-Force Kevin probe force microscopy (KPFM)}
In general, the work function value $\Phi_{sample}$ and, consequently, Fermi level variation can be calculated from KPFM measurements using an equation:
\begin{equation}
	\Phi_{sample} = e \, V_{CPD} - \Phi_{probe}	
	\label{SI-03}
\end{equation}
where $V_{CPD}$ is the contact potential difference between the sample and the AFM probe, $e$ is elemental charge, and $\Phi_{probe}$ is the work function of the KPFM probe. Prior to measuring the MoS$_2$/graphene samples, we checked the KPFM probe response against an Au-Si-Al standard. Topography and $V_{CPD}$ maps of standard are shown on Figure \ref{fig:SI-04}a,b. Next, the work function of aluminum was calibrated against a freshly cleaved highly oriented pyrolytic graphite (HOPG) reference (Figure \ref{fig:SI-04}c,d) using value of $\Phi_{HOPG} =4.6$eV (PFKPFM-SMPL, HOPG-12M, Bruker SPM Probes, Camarillo, CA). To eliminate influence of water condensation the AFM chamber was purged with nitrogen gas.

\begin{figure*}[!]
	\centering 
	\includegraphics[width=1 
	\columnwidth]{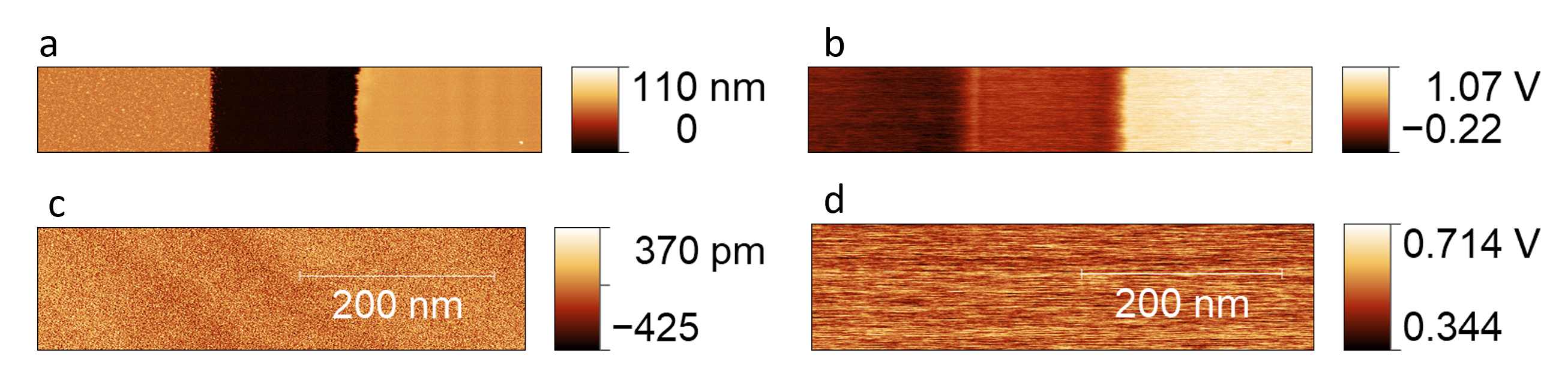}
	\caption{(a) AFM topography and (b) KPFM signal for the Au-Si-Al standard. (c) AFM topography and (d) KPFM channel for HOPG reference sample.}
	\label{fig:SI-05}
\end{figure*}

The topography of MoS$_2$/graphene sample (Figure \ref{fig:SI-05}a) is dominated by roughness of Si/SiO$_2$ substrate, and since the thickness of MLG and monolayer MoS$_2$ is below RMS, topographical details of heterostructures cannot be resolved by routine AFM imaging. The spatial distribution of $V_{CPD}$ and calculated work function value of the same area are presented on Figure \ref{fig:SI-05}b,c. It must be noted that the KPFM probe is in contact with the outermost layer of the heterostructure, graphene, thus it measures the work function of MLG, $\Phi_{MLG}$, either on or off the MoS$_2$ island. The work function value "off" the island reveals p-doping of graphene, likely due to transfer procedure. The Fermi level value, taken with respect to graphene Dirac point, is negative. The "on"  value is shifted towards the Dirac point, showing non-uniform n-doping effect, originated from the charge transfer in the heterostructure.

\begin{figure*}[!]
	\centering 
	\includegraphics[width=1 
	\columnwidth]{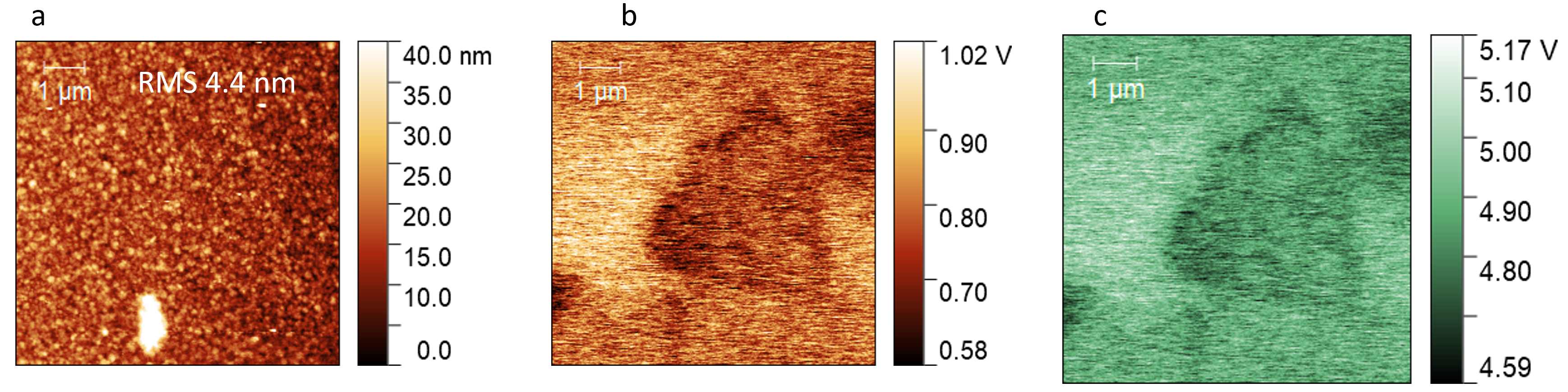}
	\caption{The maps of a MoS$_2$/graphene sample: (a) AFM topography, (b) KPFM, and (c) calculated work function distribution.}
	\label{fig:SI-06}
\end{figure*}

\section{Calculation of charge density in MLG and in  MoS$_2$ monolayer}
\label{app:dos-eqs}

For 2d-materials with parabolic dispersion relation (with massive fermions), like MoS$_2$, the energy is given by: $E=E_c+\hbar^2k^2/(2m^*)$. Then, the density of states (DOS) is constant for each band: $=2m^*/(\pi\hbar^2)$. Then, the following integral gives the carrier density dependence on the Fermi level (spin and valley degeneracy included):
\begin{equation}
	n(F)=\frac{2m^*}{\pi\hbar^2}\int_{E_c}^\infty \frac{dE}{1+\exp[\frac{E-
			F}{kT}]}=\frac{2m^* kT}{\pi\hbar^2}\log\left(1+\exp\left[\frac{F-E_c}{kT}\right]\right)=N_c \, \log\left(1+\exp\left[\frac{|E_c|-|F|}{kT}\right]\right)
	\label{2d-DOS-MoS2}
\end{equation}
where we assume that both Fermi level $F$ and $E_c=-4.21$~eV\cite{Larentis2014} are taken with respect to the vacuum level and, thus, are negative (this definition is consistent with the definition for Dirac point $E_D$, conduction band edge $E_c$ 
and Fermi level $F$
. The conduction band DOS is given by:
\begin{equation}
	N_c=\frac{2m^* kT}{\pi\hbar^2}=\frac{2m^*}{m_o}\frac{kT}{\pi a_B^2 \; E_B}\simeq 7.6\,10^{12}\,\mathrm{cm}^{-2}
	\label{Nc-MoS2}
\end{equation}
with $m_o$ being the free electron mass, $a_B=0.53$~\AA, $E_B=27$~eV, and effective mass in MoS$_2$ is taken to be 0.35$m_o$\cite{Peelaers2012}.


There are two limits to be noted: for non-degenerate doping ($|F|>|E_c|$, Fermi level lies below the bottom of CB), one can use $\log(1+x)\sim~x$ and write:
\begin{equation}
	n\simeq N_c \, \exp\left[-\frac{|F|-|E_c|}{kT}\right]
	\label{2d-DOS-MoS2-nondeg}
\end{equation}
while in the degenerate doping limit ($|E_c|-|F|\gg kT>0$, Fermi level is within the CB), unity is neglected compared to the large exponential, and we derive linear dependence of the charge density on the Fermi level:
\begin{equation}
	n\simeq N_c \frac{|E_c|-|F|}{kT} 
	\label{2d-DOS-MoS2-ndeg}
\end{equation}

Correspondingly for the monolayer graphene, which is gapless with a linear dispersion relation $E=\hbar v_F k$, we derive:
\begin{equation}
	n_g(F)=\frac{(E_D-F)^2}{\pi\hbar^2 v_F^2}=N_g\,(E_D-F)^2
	\label{2d-DOS-MLG}
\end{equation}
where the Dirac point $E_D=\chi_{MLG}\simeq -4.57$~eV\cite{Yu2009}, and Fermi velocity $v_F\simeq 1.16\times10^6$~m/s\cite{Knox2008}. We emphasize that $N_g$ is not a density of carriers, neither it is a 2d-DOS in a classical sense: $N_g\simeq 5.46\;10^{13}$~cm$^{-2}$~eV$^{-2}$.

\begin{figure*}[!]
	\centering 
	\includegraphics[width=0.5 
	\columnwidth]{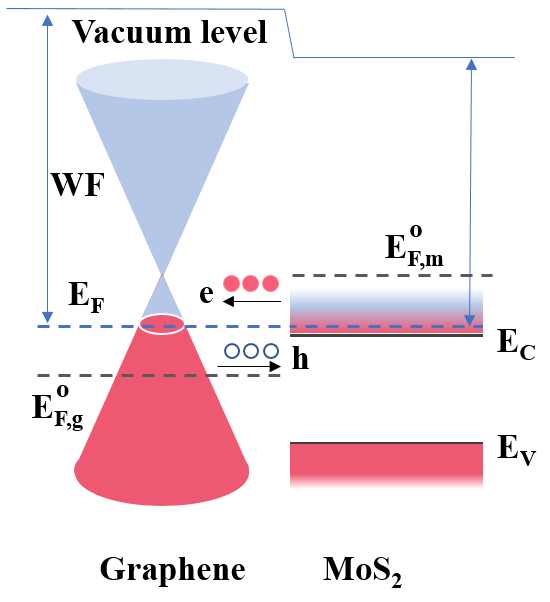}
	\caption{Matching band structure offsets in MoS$_2$/graphene van der Waals heterojunction: in order to align the Fermi level, the charg transfer between the 2D-materials must happen, resulted in a drop of vacuum level between the layers.}
	\label{fig:SI-07}
\end{figure*}

Since the 2d materials are electrically isolated from the Si substrate by the oxide layer, they are at floated potential and the charge transfer produces 2D charge densities $\pm en_1$, equal (by magnitude and opposite by sign) in both layers, and generates $2\delta V$, a potential difference between TMDC and MLG ($\phi(z\pm d/2)=\pm\delta V$). This potential difference is linearly proportional to the surface charge  formed at each of the materials, as the result of charge transfer.

Then, the positions of the Fermi levels, both defined with respect to the higher vacuum level in MLG, are:
\begin{equation}
	|F_g| =|F_g^{(o)}| -\Delta_F \qquad\qquad
	|F_{MoS_2}|=|F_{MoS_2}^{(o)}|+2\delta V +\Delta_F{}_{MoS_2}
	\label{levels-03}
\end{equation}
where the differences: $\Delta_F=F^{(o)}_g-F>0$ is the Fermi level (up)shift in graphene, which can be measured as work function difference taken on and off the TMDC island, and $\Delta_F{}_{MoS_2}$, the Fermi level (down)shift in MoS$_2$.

Knowing the expression for TMDC and MLG DOS, one can easily calculate the charge transfer and, then, the potential difference between the layers in the vertical heterostructure. Thus, the relation between the   measured MLG work function and the doping level of TMDC can be established, as shown in the Fig. 4d of main text.

\section{Animation File}
\label{app:movie}

SI includes an animation file representing some of the channels of multidimensional characterization of a particular van der Waals vertical heterostructure (single layer  MoS$_2$ packaged by graphene monolayer). The same area of the sample has been imaged using different instrumentation. Theoretical results on the charge transfer (doping) and various components of strain are also shown.

~\newpage~
\section*{References}
\label{app:ref}

\bibliography{biosens-hetero-20}



\end{document}